\documentclass[prl,twocolumn,10pt]{revtex4}%
\usepackage{amsfonts}
\usepackage{amsmath}
\usepackage{amssymb}
\usepackage{graphicx}%
\providecommand{\U}[1]{\protect\rule{.1in}{.1in}}
\begin{document}
\title{Origin of adiabatic and non-adiabatic spin transfer torques in current-driven
magnetic domain wall motion}
\author{Jun-ichiro Kishine}
\affiliation{Department of Basic Sciences, Kyushu Institute of Technology, Kitakyushu,
804-8550, Japan}
\author{A. S. Ovchinnikov}
\affiliation{Department of Physics, Ural State University, Ekaterinburg, 620083 Russia}
\date{24 November 2009}

\begin{abstract}
A consistent theory to describe the correlated dynamics of quantum mechanical
itinerant spins and semiclassical local magnetization is given. We consider
the itinerant spins as quantum mechanical operators, whereas local moments are
considered within classical Lagrangian formalism. By appropriately treating
fluctuation space spanned by basis functions, including a zero-mode wave
function, we construct coupled equations of motion for the collective
coordinate of the center-of-mass motion and the localized zero-mode coordinate
perpendicular to the domain wall plane. By solving them, we demonstrate that
the correlated dynamics is understood through a hierarchy of two time scales:
Boltzmann relaxation time $\tau_{\text{el}}$, when a non-adiabatic part of the
spin-transfer torque appears, and Gilbert damping time $\tau_{\text{DW}}$,
when adiabatic part comes up.

\end{abstract}
\maketitle

\baselineskip10.5pt

Spin torque transfer (STT) process is expected to revolutionize the
performance of memory device due to non-volatility and low-power consumption.
To promote this technology, it is essential to make clear the nature of the
current-driven domain wall (DW) motion\cite{Slonczewski,Berger}. Recent
theoretical
\cite{Bazaliy1998,Zhang-Levy-Fert2002,StilesZangwill02,Zhang2004,Tatara2004,Tatara-Shibata-Khono,Thiaville05,Xiao-Zangwill-Stiles06}
and experimental\cite{Experiments} studies have disclosed that the STT
consists of two vectors perpendicular to the local magnetization
$\boldsymbol{m}(x)$ and can be written in general as $\boldsymbol{N}=$
$c_{1}\partial_{x}\boldsymbol{m}+$ $c_{2}\boldsymbol{m}\times\partial
_{x}\boldsymbol{m}$\cite{StilesZangwill02}. The $c_{1}$ and $c_{2}$-terms
respectively come from adiabatic\cite{Slonczewski,Bazaliy1998} and
non-adiabatic \cite{Zhang2004} processes between conduction electrons and
local magnetization, and the terminal velocity of a DW is controlled by not
$c_{1}$ but small $c_{2}$ term. The origin of the $c_{2}$ term is ascribed to
the spatial mistracking of spins between conduction electrons and local
magnetization\cite{Zhang2004}. Behind appearance of the $c_{2}$ term is the so
called transverse spin accumulation (TSA)\ of itinerant electrons generated by
the electric current\cite{Zhang-Levy-Fert2002,Xiao-Zangwill-Stiles06}. Now,
any consistent theory should explain how the adiabatic and non-adiabatic STT
come up starting with microscopic model. In particular, it should be made
clear how the TSA caused by the non-adiabatic STT eventually leads to
translational motion of the whole DW. In this letter, to solve this highly
debatable problem, we propose a consistent theory to describe the correlated
dynamics of quantum mechanical itinerant spins and semiclassical local magnetization.

We consider a single head-to-head N\'{e}el DW through a magnetic nanowire with
an easy $x$ axis and a hard $z$ axis. Fee electrons travel along\ the DW axis
($x$-axis). W{e describe a} local spin by{ a semiclassical vector
}$\boldsymbol{S}${$=S\boldsymbol{n}{=}S(\sin\theta\cos\varphi,\sin\theta
\sin\varphi,\cos\theta)$ where }$S=\left\vert \boldsymbol{S}\right\vert $ and
the {polar coordinates }$\theta${\ and }$\varphi${ are assumed to be slowly
varying functions of one-dimensional coordinate }${x}${ [}Fig.1(a)]. {The DW
formation is described by the Hamiltonian (energy per unit area) in the
continuum limit,%
\begin{equation}
\mathcal{H}_{\text{DW}}=\frac{JS^{2}}{2a}{%
%TCIMACRO{\dint _{-\infty}^{\infty}}%
%BeginExpansion
{\displaystyle\int_{-\infty}^{\infty}}
%EndExpansion
}dx\left[  \left(  \partial_{x}\boldsymbol{n}\right)  ^{2}-\lambda^{-2}\hat
{n}_{x}^{2}+\kappa^{-2}\hat{n}_{z}^{2}\right]  , \label{Hamiltonian_DW}%
\end{equation}
where} $a$ is the cubic lattice constant, {$J$}\ is {the ferromagnetic
exchange strength, $\lambda=\sqrt{J/K}$ and $\kappa=\sqrt{J/K_{\bot}}$
respectively represent the single-ion easy and hard axis anisotropies measured
in the length dimension. The stationary N\'{e}el wall ($\theta_{0}=\pi/2$) is
described by $\boldsymbol{n}_{0}=(\cos\varphi_{0},\sin\varphi_{0},0)$ with
${\varphi_{0}(z)=2\arctan(e^{x/\lambda})}${. }In the infinite continuum
system, the DW{ configuration has continuous degeneracy\ labeled by }the
{center of mass position, }${X}${, of the DW. This degeneracy
apparently\ leads to \textit{rigid} translation of the DW, i.e.,
}${\boldsymbol{n}_{0}(x)\rightarrow\boldsymbol{n}_{0}(x-X)}${\cite{BM05}}}%
${.}${ As explicitly shown below, however, the translation in off-equilibrium
accompanies \textit{internal deformation} of the DW.}

\begin{figure}[b]
\begin{center}
\includegraphics[width=60mm]{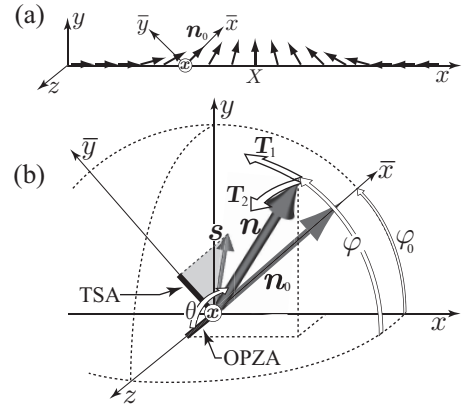}
\end{center}
\caption{(a) Stationary configuration of local spins ($\boldsymbol{n}_{0}$)
associated with a single N\'{e}el wall. Labotatory frame $x,$ $y,$ $z$ and
local frame $\bar{x},$ $\bar{y},$ $z$ are indicated. (b) Schematic view of the
transverse spin accumulation (TSA) of itinerant spin $\boldsymbol{s}$\ and the
out-of-plane ($\theta$) zero-mode accumulation (OPZA) of local spin
$\boldsymbol{n}$. These magnetic accumulations respectively cause the
non-adiabatic torque $\boldsymbol{T}_{2}$ and adiabatic torque $\boldsymbol{T}%
_{1}$. }%
\end{figure}

The creation operator of a conduction electron is written in a spinor form as
$c^{\dag}(x)=(c_{\uparrow}^{\dag}(x),c_{\downarrow}^{\dag}(x))$. By performing
the local gauge transformation $c(x)={\hat{U}}(x)\bar{c}(x)$ with the unitary
operator ${\hat{U}}(x)=e^{i{\hat{\sigma}}_{z}{\varphi_{0}(x)/2}}$
(${\hat{\sigma}}_{z}$ is a Pauli matrix) the quantization axis becomes
parallel to the local spin located at $x$. Assuming $\left\vert a{\partial
_{x}\varphi_{0}(x)}\right\vert \simeq a/\lambda\ll1$, i.e. wall thickness is
much larger than atomic lattice constant, this procedure leads to the
single-particle Hamiltonian,%
\begin{equation}
\mathcal{H}_{\text{el}}=\frac{\hbar^{2}}{2m^{\ast}a}{%
%TCIMACRO{\dint _{-\infty}^{\infty}}%
%BeginExpansion
{\displaystyle\int_{-\infty}^{\infty}}
%EndExpansion
dx}\left[  \frac{1}{2}\left\vert \partial_{x}\bar{c}\right\vert ^{2}%
+i(\partial_{x}\bar{c}^{\dag})\mathcal{{\hat{A}}}_{z}\bar{c}\right]
+\text{c.c}, \label{single-particle}%
\end{equation}
where the effective mass of the conduction electron is $m^{\ast}$. The
SU(2)\ gauge field\cite{Volovik87,Tatara2004} is introduced as $\mathcal{{\hat
{A}}}_{z}\equiv i^{-1}{\hat{U}}^{-1}\partial_{x}{\hat{U}}{=-}(\partial
_{x}{\varphi_{0}}){\hat{\sigma}}_{z}/2$. The conduction electrons are assumed
to interact with the local spins by a s-d coupling represented in the form,%
\begin{equation}
\mathcal{H}{_{\text{sd}}=-}\dfrac{{J_{sd}}}{a^{3}}{%
%TCIMACRO{\dint _{-\infty}^{\infty}}%
%BeginExpansion
{\displaystyle\int_{-\infty}^{\infty}}
%EndExpansion
dx\ \boldsymbol{\hat{s}(}x)\cdot\boldsymbol{S}{(x-X),}} \label{Hamiltonian_sd}%
\end{equation}
where ${\boldsymbol{\hat{s}}}$ and $\boldsymbol{S}=S\boldsymbol{n}$ are
respectively\ the spins of itinerant and localized electrons$.$ We
treat${\ \boldsymbol{\hat{s}(}x)=}\frac{1}{2}c^{\dag}\boldsymbol{\hat{\sigma}%
}c$ as fully quantum mechanical operator, while $\boldsymbol{n}$ is a
semiclassical vector.

\textit{Boltzmann relaxation}: l{et switch on the electric field $E$\ at
$t=0$. We introduce the Boltzmann relaxation time $\tau_{\text{el}}$\ and }the
number density of the conduction electrons $f_{k\sigma}$ in the state
$k,\sigma$. We assume that the deviation from equilibrium Fermi-Dirac
distribution $f_{0}\left(  \varepsilon_{k\sigma}\right)  =\left[  \exp\left[
\left(  \varepsilon_{k\sigma}-\mu\right)  /k_{B}T\right]  +1\right]  ^{-1}$ is
small, where $\varepsilon_{k\sigma}$\ ($\sigma=\uparrow,\downarrow$)\ is the
single-particle energy, $\mu$\ is the chemical potential. Using standard
Boltzmann kinetic equation with relaxation time
approximation\cite{Xiao-Zangwill-Stiles06}, the distribution function is
written as%
\begin{equation}
f_{k\sigma}\simeq f_{0}\left(  \varepsilon_{k\sigma}\right)  +eE\tau
_{\text{el}}v_{k\sigma}\frac{\partial f_{0}\left(  \varepsilon_{k\sigma
}\right)  }{\partial\varepsilon_{k\sigma}}, \label{Distribution_fun}%
\end{equation}
where the electron charge is $-e$ and the spin-dependent velocity is
$v_{k\sigma}\equiv\hbar^{-1}\partial\varepsilon_{k\sigma}/\partial k$. The
spin-dependence of $\varepsilon_{k\sigma}$ originates from the SU(2) gauge
fields $(\mathcal{{\hat{A}}}_{z})_{\uparrow\uparrow}$ and $(\mathcal{{\hat{A}%
}}_{z})_{\downarrow\downarrow}$. In the process of approaching to stationary
current flowing state around the time $t\sim\tau_{\text{el}}$, as we will show
explicitly, the statistical average of the conduction electron's spin
component perpendicular to the local quantization axis accumulates and
acquires finite value. As schematically depicted in Fig.1(b), this process is
exactly the TSA. The TSA causes an additional magnetic field acting on the
local spins and exert the non-adiabatic torque on the local spins.

\textit{Local spin dynamics}: next we formulate dynamics of the local spins
coupled with the conduction electrons. We introduce the {$\delta\theta\left(
x,t\right)  $ (out-of-plane) and $\delta\varphi\left(  x,t\right)  $
(in-plane)\ fluctuations of the local spins around the stationary DW
configuration }${\boldsymbol{n}_{0}(x)}${$.$ We say \textquotedblleft%
{out-of-plane\textquotedblright} and \textquotedblleft
in-plane\textquotedblright\ with respect to the DW plane. The fluctuations are
spanned by the }orthogonal{ basis functions }$v_{q}\ ${and }$u_{q}${\ as
$\varphi\left(  x\right)  =\varphi_{0}\left(  x-X\right)  +\delta
\varphi\left(  x-X\right)  $ and $\theta\left(  x\right)  =\pi/2+\delta
\theta\left(  x-X\right)  ,$ where}%
\begin{equation}
\delta\varphi\left(  x\right)  ={%
%TCIMACRO{\dint _{-\infty}^{\infty}}%
%BeginExpansion
{\displaystyle\int_{-\infty}^{\infty}}
%EndExpansion
dq}\text{ }{\eta_{q}(t)v_{q}\left(  x\right)  ,}\text{ }{\delta\theta\left(
x\right)  }{=%
%TCIMACRO{\dint _{-\infty}^{\infty}}%
%BeginExpansion
{\displaystyle\int_{-\infty}^{\infty}}
%EndExpansion
{dq}}\text{ }{\xi_{q}(t)u_{q}\left(  x\right)  .} \label{v}%
\end{equation}
{At this stage, }${X}${\ is not a dynamical variable, but just a parameter.
The basis functions obey the Schr\"{o}dinger equations, }$\left(
JS^{2}/2\right)  $ $(-\partial_{x}^{2}$ $-2\lambda^{-2}\sin^{2}\varphi_{0}$
$+\lambda^{-2})$ $v_{q}\left(  x\right)  $ $=$ $\varepsilon_{q}^{\varphi}%
u_{q}\left(  x\right)  $ and $\left(  JS^{2}/2\right)  $ $(-\partial_{x}^{2}$
$-2\lambda^{-2}\sin^{2}\varphi_{0}$ $+\lambda^{-2}$ $+\kappa^{-2})$
$u_{q}\left(  x\right)  $ $=$ $\varepsilon_{q}^{\theta}u_{q}\left(  x\right)
.${\ Both $\theta$ and $\varphi$ modes consist of a single bound state
(\textit{zero mode})\ and continuum states (\textit{spin-wave modes}). The
dimensionless zero mode wave functions are given by }${u_{0}\left(  x\right)
=v_{0}\left(  x\right)  ={\Phi}_{0}(x)}$, where%
\begin{equation}
{{\Phi}_{0}(x)\equiv}\sqrt{\dfrac{a\lambda}{2}}\partial_{x}{\varphi_{0}%
(x)=}\sqrt{\dfrac{a}{2\lambda}}\dfrac{1}{\cosh(x/\lambda),}
\label{zero-mode-wave-fun}%
\end{equation}
{with the corresponding energies respectively given by }$\varepsilon
_{0}^{\theta}=JS^{2}/(2\kappa^{2})${\ and }$\varepsilon_{0}^{\varphi}=0${. The
normalization is given by }${a}^{-1}{%
%TCIMACRO{\tint _{-\infty}^{\infty}}%
%BeginExpansion
{\textstyle\int_{-\infty}^{\infty}}
%EndExpansion
dx}\left[  {{\Phi}_{0}(x)}\right]  ^{2}=1.${ {Although to excite the
out-of-plane ($\theta$)\ zero mode costs finite energy gap }}$\varepsilon
_{0}^{\theta}$ coming from the hard-axis anisotropy{{, we still call this
\textquotedblleft zero mode.\textquotedblright\ }The spin-wave states have
energy dispersions given by $\varepsilon_{q}^{\theta}=\tfrac{1}{2}JS^{2}$%
}$\left(  {q^{2}+\lambda^{-2}+\kappa^{-2}}\right)  ${ and $\varepsilon
_{q}^{\varphi}={\tfrac{1}{2}JS^{2}}$}$\left(  {q^{2}+\lambda^{-2}}\right)  ${.
Because the zero mode and the spin-wave states are orthogonal to each other
and separated by the anisotropy gaps, the spin-wave modes are totally
irrelevant to a low energy effective theory. Therefore, we ignore the
spin-wave modes from now on.}

\textit{Out-of-plane zero-mode(OPZ) coordinate }$\xi_{0}$: {in order to obtain
the correct form of the dynamical Hamiltonian, one has to regard the variable
$X$ as a dynamical variable $X(t)$ and replace the zero mode coordinate
$\eta_{0}$ with $X(t)$. Following this idea, the zero-mode fluctuations should
be given by, }
\begin{align}
\varphi(x,t)  &  =\varphi_{0}\left[  x-X\left(  t\right)  \right]
,\label{Collectivephi}\\
\theta(x,t)  &  =\pi/2+\xi_{0}(t){\Phi}_{0}\left[  x-X\left(  t\right)
\right]  . \label{Collectivetheta}%
\end{align}
{ Eq. (\ref{Collectivetheta}) is a key ingredient of this letter, which has
never been explicitly treated so far\cite{BKO08}. That is to say, we naturally
include the }out-of-plane(OPZ) zero-mode, in addition to the in-plane
($\varphi$) zero-mode replaced by $X\left(  t\right)  ${. }The zero-mode wave
function $\Phi_{0}\left[  x-X\left(  t\right)  \right]  $\ serves as the basis
function of the $\theta$-fluctuations localized around the center of the DW
and $\xi_{0}(t)$\ is the OPZ \textit{coordinate}. Now, our effective theory is
fully described by {two dynamical variables }${X(t)}$ and $\xi_{0}(t)${ which
naturally give physical coordinates along the Hilbert space of orthogonal\ }%
$\theta${\ and }$\varphi${\ fluctuations. As we will see, we have }$\xi
_{0}(t)\neq0$ only for inequilibrium current flowing state under $E\neq0$
[Fig. 2(a)].

\begin{figure}[t]
\begin{center}
\includegraphics[width=65mm]{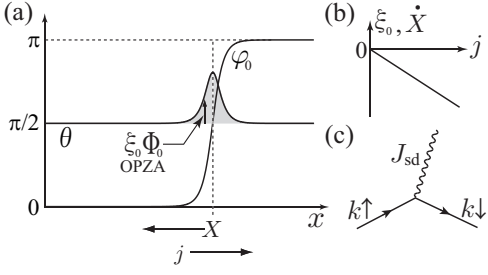}
\end{center}
\caption{(a) Spatial profile of the polar angles $\varphi(x,t)=\varphi
_{0}[x-X\left(  t\right)  ]$ and $\theta(x,t)=\pi/2+\xi_{0}(t){\Phi}%
_{0}\left[  x-X\left(  t\right)  \right]  $ in the current flowing state. (b)
Linear dependence of $\xi_{0}$ and $\dot{X}\left(  t\right)  $ on the current
density $j$. (c) Single- particle propagation (represented by solid line) with
spin flip process by the s-d interaction (represented by wavy line) which
leads to the STT.}%
\label{g2}%
\end{figure}

{It is here important to note an essential difference between Tatara and
Khono's approach}\cite{Tatara2004} and ours. Tatara and Khono used $X(t)$ and
the weighted average, $\theta_{0}(t)$ $=$ ${%
%TCIMACRO{\tint _{-\infty}^{\infty}}%
%BeginExpansion
{\textstyle\int_{-\infty}^{\infty}}
%EndExpansion
}dx$ $\theta(x,t)\sin^{2}\varphi\lbrack x-X(t)]$, as dynamical variables.
Later, they systematically used complex coordinate $\xi=e^{i\varphi}%
\tan(\theta/2)$ and described the fluctuations in the form $\xi
=e^{-u(x,t)+i\varphi_{0}+\eta\lbrack x-X(t)]}$\cite{Tatara-Shibata-Khono}%
(their notation is reproduced by putting $\theta\rightarrow\pi/2-\theta$,
$\varphi\rightarrow\varphi$ in our notation). In our understanding, these
descriptions inevitably cause redundant coupling between $u$ and $v$ modes in
Eq. (\ref{v}). Actually, our natural choice of the dynamical variables is
essential to appropriately derive relaxational dynamics described by the
following equations of motion given by {(\ref{EOM1}) and (\ref{EOM2})}.

\textit{Equations of motion of the DW}: {now, we construct an effective
Lagrangian }$\mathcal{L}$$=\mathcal{L}_{\text{DW}}+\mathcal{L}_{\text{sd}}%
${\ to describe the DW motion and resultant equations of motion (EOM). Using
(\ref{Collectivephi}) and (\ref{Collectivetheta}), the local spin counterpart
is given by }$\mathcal{L}_{\text{DW}}=\tfrac{{\hbar S}}{a^{3}}{%
%TCIMACRO{\tint _{-\infty}^{\infty}}%
%BeginExpansion
{\textstyle\int_{-\infty}^{\infty}}
%EndExpansion
dx\left(  \cos\theta-1\right)  \dot{\varphi}-}\mathcal{H}{_{\text{DW}}}$
explicitly written as%
\begin{equation}
\mathcal{L}_{\text{DW}}=\dfrac{{\hbar S}}{a^{3}}\left(  \sqrt{\dfrac
{2a}{\lambda}}\xi_{0}+\pi\right)  \dot{X}-\dfrac{JS^{2}}{2\kappa^{2}}\xi
_{0}^{2}. \label{Lagrangian_DW}%
\end{equation}
To understand the effect of the s-d coupling, it is useful to note
$\boldsymbol{n}{\left[  \theta_{0}+\delta\theta,\varphi_{0}+\delta
\varphi\right]  }$ $\simeq$ ${{\boldsymbol{n}_{0}}}$ ${{-\boldsymbol{e}%
_{z}\delta\theta}}$ ${-{\boldsymbol{n}_{0}}\delta\theta^{2}/2,}$ where {we
dropped }$\delta\varphi$\ because this degree of freedom is eliminated by the
global gauge fixing\cite{BKO08}. We have thus s-d Lagrangian,%
\begin{equation}
\mathcal{L}_{\text{sd}}=a^{-3}J_{sd}S\left(  \mathcal{F}_{0}-\mathcal{S}%
_{\Vert}\xi_{0}{}^{2}/2\right)  , \label{Lagrangian_sd}%
\end{equation}
{where,} $\mathcal{F}_{0}\left[  X\left(  t\right)  \right]  $ $\equiv{%
%TCIMACRO{\tint _{-\infty}^{\infty}}%
%BeginExpansion
{\textstyle\int_{-\infty}^{\infty}}
%EndExpansion
}dx$ $\boldsymbol{\hat{n}}_{0}\left[  x-X\left(  t\right)  \right]  \cdot$
$\left\langle \boldsymbol{s}(x,t)\right\rangle $ and $\mathcal{S}_{\Vert
}\left[  X\left(  t\right)  \right]  $ $\equiv$ ${%
%TCIMACRO{\tint _{-\infty}^{\infty}}%
%BeginExpansion
{\textstyle\int_{-\infty}^{\infty}}
%EndExpansion
}dx$ $\left\{  {\Phi}_{0}\left[  x-X\left(  t\right)  \right]  \right\}
^{2}\boldsymbol{n}_{0}\left[  x-X\left(  t\right)  \right]  \cdot$
$\left\langle \boldsymbol{s}(x,t)\right\rangle $. {Finally, to take account of
dissipative dynamics, we use the Rayleigh dissipation function} $\mathcal{W}%
{_{\text{Rayleigh}}=}\tfrac{\alpha}{2}\tfrac{{\hbar S}}{a^{3}}{%
%TCIMACRO{\tint _{-\infty}^{\infty}}%
%BeginExpansion
{\textstyle\int_{-\infty}^{\infty}}
%EndExpansion
dx}$ ${\boldsymbol{\dot{n}}^{2}}$ explicitly written as%
\begin{equation}
\mathcal{W}{_{\text{Rayleigh}}}=\dfrac{\alpha}{2}\dfrac{{\hbar S}}{a^{3}%
}\left(  a\dot{\xi}_{0}^{2}+\dfrac{2}{\lambda}\dot{X}^{2}\right)  ,
\label{Rayleigh}%
\end{equation}
where $\alpha$\ is the Gilbert damping parameter. It is simple to write down
{the Euler-Lagrange-Rayleigh equations, $d\left(  \partial L/\partial\dot
{q}_{i}\right)  /dt$ $-$ $\partial L/\partial q_{i}$ $=$ $-\partial
W/\partial\dot{q}_{i},$ for the dynamical variables $q_{1}=X$ and $q_{2}%
=\xi_{0}$. We obtain the EOMs }which contain the dynamical variables in linear
order,
\begin{subequations}
\begin{gather}
{\hbar}\sqrt{\dfrac{2a}{\lambda}}\dot{\xi}_{0}+J_{sd}\mathcal{T}_{\bot
}=-2\alpha\dfrac{{\hbar}}{\lambda}\dot{X},\label{EOM1}\\
-{\hbar}\sqrt{\dfrac{2a}{\lambda}}\dot{X}+\left(  \dfrac{a^{3}JS}{\kappa^{2}%
}+J_{sd}\mathcal{S}_{\Vert}\right)  \xi_{0}=-\alpha{\hbar}a\dot{\xi}_{0},
\label{EOM2}%
\end{gather}
where the quantities
\end{subequations}
\begin{subequations}
\begin{align}
\mathcal{T}_{\bot}  &  \equiv-\frac{\partial\mathcal{F}_{0}}{\partial X}={%
%TCIMACRO{\dint _{-\infty}^{\infty}}%
%BeginExpansion
{\displaystyle\int_{-\infty}^{\infty}}
%EndExpansion
}dx\text{ }\partial_{x}\varphi_{0}\left[  x-X\left(  t\right)  \right]
\left\langle \bar{s}_{y}\boldsymbol{(}x\boldsymbol{)}\right\rangle
,\label{spin_torque}\\
\mathcal{S}_{\Vert}  &  \equiv{%
%TCIMACRO{\dint _{-\infty}^{\infty}}%
%BeginExpansion
{\displaystyle\int_{-\infty}^{\infty}}
%EndExpansion
}dx\text{ }{\Phi}_{0}^{2}\left[  x-X\left(  t\right)  \right]  \left\langle
\bar{s}_{x}\boldsymbol{(}x\boldsymbol{)}\right\rangle ,
\label{spin_accumulation}%
\end{align}
respectively give the non-adiabatic STT and longitudinal spin
accumulation\cite{Zhang-Levy-Fert2002}. The statistical average of the
conduction electron's spin component is denoted by $\left\langle
\cdots\right\rangle $. The gauge-transformed spin variables are introduced by
$\boldsymbol{\bar{s}}$$(x)={\hat{U}}^{-1}\left[  {x-X(t)}\right]  $
${\boldsymbol{\hat{s}}(x)}{\hat{U}}\left[  {x-X(t)}\right]  $ which
has{\ local quantization axis tied to the local spin at the position of
}${x-X(t)}${. To obtain Eq.(\ref{spin_torque}), we used relations }%
$\partial_{x}\mathbf{n}_{0}\left[  {x-X(t)}\right]  =-\partial_{x}\varphi
_{0}(x)$ $\mathbf{e}_{z}\times\mathbf{n}_{0}\left[  {x-X(t)}\right]  $ and
$\left\langle \bar{s}_{y}\right\rangle =-\left\langle \hat{s}_{x}\right\rangle
\sin\varphi_{0}+\left\langle \hat{s}_{y}\right\rangle \cos\varphi_{0}.$ The
relation (\ref{spin_torque}) implies that the translation of the DW
($x\rightarrow x-X$)\ naturally gives rise to the TSA, $\left\langle \bar
{s}_{y}\right\rangle $,\ along the local $\bar{y}$ axis. The appearance of
$\left\langle \bar{s}_{y}\right\rangle $ causes local magnetic moment which
triggers the local spins to precess around the local $\bar{y}$ axis and
consequently produce finite deviation of the polar angle $\delta\theta
=\theta-\theta_{0}.$ It is seen that upon switching the external electric
field, the deviation $\delta\theta$\ relaxes to finite magnitude in the
stationary current-flowing state, i.e., the OPZ coordinate $\xi_{0}(t)$
accumulates and reaches finite terminal value $\xi_{0}^{\ast}$. We call this
process out-of-plane zero-mode accumulation(OPZA)as schematically depicted in
Fig.1(b){. }This effect is physically interpreted as appearance of
demagnetization field phenomenologically introduced by D\"{o}ring, Kittel,
Becker\cite{Doring48}, and Slonczewski\cite{Slonczewski}. It is also to be
noted that we ignored the term $\partial\mathcal{S}_{\Vert}/\partial X$. This
simplification is legitimate for the case of of small sd-coupling.

\textit{Gilbert}{ }\textit{relaxation}: {coupled equations of motion
(\ref{EOM1}) and (\ref{EOM2}) are readily solved to give relaxational
solutions,}%
\end{subequations}
\begin{equation}
\xi_{0}=\xi_{0}^{\ast}(1-e^{-t/\tau_{\text{DW}}}),\text{ \ \ }V\equiv\dot
{X}=V^{\ast}(1-e^{-t/\tau_{\text{DW}}}), \label{terminal}%
\end{equation}
where {the OPZA reaches the }terminal value,%
\begin{equation}
\xi_{0}^{\ast}=-\frac{1}{\alpha}\dfrac{\sqrt{a\lambda/2}J_{sd}\mathcal{T}%
_{\bot}}{\left(  a^{3}JS\kappa^{-2}+J_{sd}\mathcal{S}_{\Vert}\right)  }%
\simeq-\alpha^{-1}\sqrt{\dfrac{\lambda}{2a}}\left(  \dfrac{\kappa}{a}\right)
^{2}\dfrac{J_{sd}}{JS}\mathcal{T}_{\bot}, \label{LZA}%
\end{equation}
and correspondingly the terminal velocity of the DW reaches $V^{\ast}%
=-\dfrac{\lambda}{2\alpha{\hbar}}J_{sd}\mathcal{T}_{\bot}.$ The relaxation
time of the DW magnetization, $\tau_{\text{DW}}$, is given by%
\begin{equation}
\tau_{\text{DW}}={\hbar}a\dfrac{\alpha^{-1}+\alpha}{\kappa^{-2}a^{3}%
JS+J_{sd}\mathcal{S}_{\Vert}}\simeq\alpha^{-1}\left(  \dfrac{\kappa}%
{a}\right)  ^{2}\dfrac{{\hbar}}{JS}. \label{rt}%
\end{equation}
This result clearly shows that {the DW magnetization try to relax through the
Gilbert damping\ toward the direction of the newly established precession
axis. We stress that without the OPZ coordinate }$\xi_{0}$ in Eqs.
(\ref{EOM1}) and (\ref{EOM2}), only the terminal velocity is available and the
transient relaxational dynamics is totally lost.

As depicted in Fig.2(a), the OPZA [Eq. (\ref{LZA})] gives rise to finite
out-of-plane ($z$) component of the local spin,%
\begin{equation}
n_{z}(x,t)=\cos\theta\simeq\frac{1}{2\alpha}\left(  \dfrac{\kappa}{a}\right)
^{2}\dfrac{J_{sd}}{JS}\dfrac{1}{\cosh\left[  (x-X(t))/\lambda\right]
}\mathcal{T}_{\bot}.
\end{equation}
The resultant local spin $\boldsymbol{S}_{\bot}=S\boldsymbol{e}_{z}n_{z}(x,t)$
gives the demagnetization field\ phenomenologically treated by Slonczewski and
gives rise to the adiabatic torque $\boldsymbol{T}_{1}=$ $c_{1}\partial
_{x}\boldsymbol{n}(x)=c_{1}\left(  \partial_{x}\varphi_{0}\right)
(-\sin\varphi_{0},\cos\varphi_{0},0)$. At the interface of the DW boundary,
$\varphi_{0}=\pi/2$ and $\boldsymbol{T}_{1}=c_{1}\left(  \partial_{x}%
\varphi_{0}\right)  (-1,0,0)$, i.e., the adiabatic torque rotate the local
spin to counterclockwise direction when the electric current flows in the
$(1,0,0)$-direction. As is clear from the above discussion, this adiabatic
torque is established \textit{after} the stationary current-flowing
[$j=(ne^{2}\tau_{\text{el}}/m^{\ast})E$] state establishes the non-adiabatic
torque, $\mathcal{T}_{\bot}$. {Around the time scale of $t\simeq
\tau_{\text{el}}+$}$\tau_{\text{DW}}${, the whole system (including conduction
electrons and DW) reaches \textit{non-equilibrium\ but stationary} state. In
this state, the DW magnetizations continuously feel the OPZA and
macroscopically rotate around it. This process exactly corresponds to
stationary translation of the DW. }

\textit{Computation of }$\mathcal{T}_{\bot}$: the final step is to compute an
explicit form of $\mathcal{T}_{\bot}$.{\ By taking Fourier transform }$\bar
{c}_{k\sigma}(t)$ $=$ $\frac{1}{\sqrt{L}}%
%TCIMACRO{\tsum _{k}}%
%BeginExpansion
{\textstyle\sum_{k}}
%EndExpansion
e^{ikx}\bar{c}_{\sigma}(x,t)$, and retaining only the momentum conserving
process{, we have}%
\begin{subequations}
\begin{align}
{\mathcal{T}_{\bot}}  &  {=}\frac{{1}}{2}{%
%TCIMACRO{\dint _{-\pi/a}^{\pi/a}}%
%BeginExpansion
{\displaystyle\int_{-\pi/a}^{\pi/a}}
%EndExpansion
dk\ \operatorname{Re}G_{k\uparrow,k\downarrow}^{<}(t,t),}\label{STT}\\
\mathcal{S}_{\Vert}  &  {=}\frac{a}{2\pi}{%
%TCIMACRO{\dint _{-\pi/a}^{\pi/a}}%
%BeginExpansion
{\displaystyle\int_{-\pi/a}^{\pi/a}}
%EndExpansion
dk\ \operatorname{Im}G_{k\uparrow,k\downarrow}^{<}(t,t).}%
\end{align}
Here, {the expectation values\ are computed by using the lesser component of
the path-oriented Green function }$G_{k\sigma,k^{\prime}\sigma^{\prime}}%
^{<}(t,t^{\prime})=i\langle\bar{c}_{k^{\prime}\sigma^{\prime}}^{\dag
}(t^{\prime})\bar{c}_{k\sigma}(t)\rangle$, where $t$ ($t^{\prime}$){ is
defined on the upper (lower) branch of Keldysh contour. }Since $\mathcal{S}%
_{\Vert}$\ does not play an essential role, we pay attention to an essential
quantity {$\mathcal{T}_{\bot}$}. To evaluate the Green functions, we
perturbatively treat the s-d coupling and write down the Dyson equation. Then,
we truncate the Dyson equation by using the Born approximation including the
s-d coupling in linear order which causes a single spin flip process
[Fig.2(c)] and gives rise to off-diagonal component in spin space,%
\end{subequations}
\begin{equation}
G_{k\uparrow,k\downarrow}^{<}(t,t)=-i\dfrac{J_{sd}}{2}\dfrac{f_{k\uparrow
}-f_{k\downarrow}}{\varepsilon_{k\uparrow}-\varepsilon_{k\downarrow}-i0}.
\label{GF}%
\end{equation}
To obtain the explicit form of $\varepsilon_{k\sigma}$, we write{\ the
single-particle Hamiltonian (\ref{single-particle}) in Fourier space and
obtain }$\mathcal{H}_{\text{el}}=\mathcal{H}_{\text{0}}+\mathcal{H}%
_{\text{gauge}}$, where $\mathcal{H}_{\text{0}}$ represents free conduction
and $\mathcal{H}_{\text{gauge}}$ comes form the second term in Eq.
(\ref{single-particle}). By retaining only momentum conserving process, we
have \ $\mathcal{H}_{\text{el}}=$ $\sum_{k,\sigma}\varepsilon_{k\sigma}\bar
{c}_{k\sigma}^{\dag}\bar{c}_{k\sigma},$ where $\varepsilon_{k\uparrow
,\downarrow}=\hbar^{2}(k\mp\delta k)^{2}/2m^{\ast}$, where the shift of the
Fermi wave numbers due to the background DW is given by $\delta k=\pi/\left(
2a\right)  $.

Using Eqs. (\ref{Distribution_fun}), (\ref{GF}), and (\ref{STT}), we finally
obtain the STT which points in the $z$-direction, $\boldsymbol{T}%
_{1}=\mathcal{T}_{\bot}\mathrm{e}_{z}$, where its magnitude is given in a
form,
\begin{equation}
\mathcal{T}_{\bot}=\frac{{1}}{4}\frac{J_{sd}}{k_{B}T}\dfrac{1}{\cosh
^{2}\left[  \left(  \varepsilon_{0}-\mu\right)  /2k_{B}T\right]  }\frac
{j}{j_{0}},\label{T}%
\end{equation}
where $j_{0}=4ne\hbar/\left(  \pi am^{\ast}\right)  $ and $\varepsilon
_{0}=\hbar^{2}\pi^{2}/(8m^{\ast}a^{2})$ corresponds to the chemical potential
at half-filling. We have a master formula which gives relation between the
current density and the terminal velocity of the DW,%
\begin{equation}
V^{\ast}=-\frac{1}{8\alpha}\dfrac{\lambda J_{sd}}{{\hbar}}\frac{J_{sd}}%
{k_{B}T}\dfrac{1}{\cosh^{2}\left[  \left(  \varepsilon_{0}-\mu\right)
/2k_{B}T\right]  }\frac{j}{j_{0}}.\label{V}%
\end{equation}
As shown in Fig.2(b), we see there is no threshold for the velocity, which is
consistent with the result obtained by Thiaville et al.\cite{Thiaville05}.
Standard choice of parameters, $j_{0}\simeq10^{16}[$A$\cdot$m$^{-2}]$,
$\lambda=10^{-8}[$m$]$, $\alpha=10^{-2}$, $j\simeq10^{11}[$A$\cdot$m$^{-2}]$
give a rough estimate $V^{\ast}\simeq-100\left(  J_{sd}/k_{B}T\right)  ^{2}%
[$m/s$].$ Of course, to pursuit more quantitative result needs numerical
estimation of $\mathcal{T}_{\bot}$ taking account of real band structure.

It is essential that the Gilbert damping coefficient, $\alpha$, enters
Eq.(\ref{V}).\ The relaxation process of the DW dynamics is governed by the
Boltzmann relaxation followed by the Gilbert damping in hierarchical manner.
As summarized in Figs.2(a) and (b), in our treatment, it is crucial to
recognize that {the OPZ coordinate }$\xi_{0\text{ }}$acquires finite value
(i.e., accumulation) only for the current flowing state which is
non-equilibrium\ but stationary{. }\textit{This is the case where dynamical
relaxation leads to finite accumulation of physical quantities which are zero
in equilibrium.} {Although }essential role of the sliding mode to describe
localized spin dynamics was pointed out before\cite{Tatara2004,BM05} and
importance of out-of-plane canting of the local spins was
stressed\cite{Slonczewski,Tatara2004}, the OPZA presented in this letter has
not been discussed before. For example, the sliding motion in Ref.\cite{BM05}
does not contain internal deformation{ of the DW. The OPZA is an outcome of
time-reversal-symmetry breaking by electric current. This interpretation seems
natural because current-flowing state is off equilibrium. }

\begin{acknowledgments}
J.~K. acknowledges Grant-in-Aid for Scientific Research (C) (No.~19540371)
from the Ministry of Education, Culture, Sports, Science and Technology, Japan.
\end{acknowledgments}

\end{document}